\documentclass[acmsmall, screen, natbib=false]{acmart}
\usepackage{fancyvrb}
\usepackage{listings}
\usepackage{mathpartir, mycommands, proof}
\usepackage{amsthm}

 \hypersetup{
           breaklinks=true,   
           colorlinks=true,   
           pdfusetitle=true,  
        }

\RequirePackage{biblatex}

\newtheoremstyle{slanted}
  {\topsep}
  {\topsep}
  {\slshape}
  {}
  {\bfseries}
  {.}
  {0.5em}
  {}

\theoremstyle{slanted}

\lstdefinestyle{style1}{
  basicstyle=\ttfamily\footnotesize
}
\begin{document}

\acmConference[SPLASH Companion '24]{Companion Proceedings of the 2024 ACM SIGPLAN International Conference on Systems, Programming, Languages, and Applications: Software for Humanity}{October 20-25}{Pasadena, California, United States}
\acmBooktitle{Companion Proceedings of the 2024 ACM SIGPLAN International Conference on Systems, Programming, Languages, and Applications: Software for Humanity (SPLASH Companion '24), October 20-25, 2024, Pasadena, California, United States}

\setcopyright{acmlicensed}

\copyrightyear{2024}

\title{Meerkat: A Distributed Reactive Programming Language with Live Updates}

\author{Anrui Liu}
\email{anruil@andrew.cmu.edu}
\affiliation{
  \institution{Carnegie Mellon University}
  \department{School of Computer Science}
  \city{Pittsburgh}
  \state{Pennsylvania}
  \country{USA}
}
\authornote{Both authors contributed equally to the research}

\author{Heng Zhong}
\email{hzhong21@m.fudan.edu.cn}
\affiliation{
  \institution{Fudan University}
  \department{School of Computer Science}
  \city{Shanghai}
  \country{P. R. China}
}
\authornotemark[1]

\begin{abstract}
  We propose a novel type-safe reactive programming language with live
  updates that extends an existing work to support multiple distributed evolution
  queues. Dependency sets of definitions are incorporated
  in the type system to protect the interaction between the frontend user
  interface and the backend database. Distributed live updates submitted by
  multiple programmers are ensured strong consistency based on an existing
  framework for distributed reactive propagation.
\end{abstract}

\maketitle

\section{Introduction} \label{sec:intro}
Reactive programming is a programming paradigm in which a 
set of persistent {\it state variables} and a set of
{\it (reactive) definitions} are defined. A state variable is
independent of any other definitions, while a
definition has a dependency set recording the state variables and
other definition that it is defined upon. An {\it action}
imperatively modifies state variables and such changes are propagated through
all subsequent definitions. An example is shown in Listing
\ref{lst:react}.
\begin{lstlisting}[caption=Reactive program example,captionpos=b,
                   label={lst:react},xleftmargin=.15\textwidth]
var x = 1;
def inc1 = x + 1;
def inc2 = inc1 + 1;
\end{lstlisting}
When {\tt x} is mutated to 2, {\tt inc1} is automatically updated to 3 and
{\tt inc2} to 4.

Current reactive programming frameworks are limited to local applications,
leaving the interaction between the client end and the server end unprotected
under update protocols. Meerkat unifies such interaction into a reactive
framework protected under a type system with explicit dependencies and enables continuous evolvement\cite{Hicks} and data-reconfiguration\cite{Bhattacharya}.

Meerkat extends a previous work by Costa Seco \cite{Seco}, which is based on a
centralized server, single programmer (conducting updates) setting. We extend
this work to support type-safe parallel submission of evolutions by multiple
programmers. Dependency graphs are stored across the frontend and backend
since they are defined in a unified application program. We leverage Historiographer \cite{Freeman} to ensure strong consistency and glitch
freedom of the propagation through the distributed dependency graph.

Section \ref{sec:dscr} presents the description and syntax of our language. Section
\ref{sec:formal} formally defines semantics of
Meerkat that secures parallel updates with the type system and the execution of concurrent actions without race condition.

\section{Language Overview and Syntax} \label{sec:dscr}
We proposed a new language Meerkat designed for 
distributed reactive programming. 
The syntax of Meerkat includes basic expressions, additionally 
declaration of state and definition variables, and actions.
\[\small
\begin{array}{lcl}
 e & ::=
 & v
 \mid x
 \mid \lambda x.e 
 \mid e_1 \, e_2
 \mid e_1 \, \mathsf{op} \, e_2 
 \mid \kwifthenelse{e}{e_1}{e_2}
 \\
 & &  \mid f \mid \kwaction{a} \\

 r & ::= & 
    \epsilon 
    \mid r \oast \kwdefvar{f}{e}
    \mid r \oast  \kwstatevar{f}{e} \\

 a & ::= & \epsilon 
 \mid  a \oast ( f := e ) \\
 
\end{array}
\]

Besides the lambda calculus core, expressions also include: (i) field names 
of state variables and definition, as a shorthand for retrieving the value 
stored in the mutable name cell; (ii) actions, behaving like lambda expressions as a sequence of writes to state variables suspended until users trigger them through ``do'' statements.

\(r\) is a sequence of declarations of definition or state variables,
representing the top most level of the program. Notice an action as expression
can be assigned to \(f\), or expression depending on other definition or state variables.
\(a\) is defined in a transactional style (by atomic composition), which means 
no partial execution of actions. Similarly, so does \(r\) since new \(r\) may 
evolve previous typing environments, and we only permit new \(r\)'s that still
generates a sound typing environments. In both cases, no partial composition will 
be granted by the system.

\section{Semantics Specification} \label{sec:formal}
The statics for Meerkat ensure that the program is well-typed, and the dependency 
graph among variables is acyclic and consistent across the distributed system.
Assuming we can distinguish state variables and definitions, we follow similar idea \cite{Seco} to maintain a typing environment \(\Delta 
    ::= \epsilon
    \mid \Delta, f \sim (\Box, \tau)
    \mid \Delta, f \sim (\delta, \tau) 
    \)
where \(\delta 
    ::= \epsilon 
        \mid \delta, f : {\tau}
    \) 
representing direct dependencies that \(f\) reads, and \(\omega\) as the set of 
variables an action writes. Types and their inference are defined w.r.t. \(\delta\) as $\tau ::= \text{basic types }
    \mid \typefunc{\tau_1}{\tau_2}{\delta}
    \mid \typeaction{\delta}{\omega}$.

Based on \(\Delta\), it is easy to statically estimate the write and (transitively) 
read sets of a variable. Write and read sets are essential for judgement of well-formness of \(\Delta\) and correct stepping of Historiographer.
We annotate function types and action types with \(\delta\) 
(and \(\omega\)) due to that both are suspended computation until applied or done when we 
actually union the read/write set.
\begin{mathpar}\small
\inferrule[]{ 
\typeinfer{\Delta; \Gamma}{\kwaction{a} : \typeaction{\delta}{\omega}} {\emptyset}{} \;\;\;
\Delta(f) \sim (\Box, \tau) \;\;\;
\typeinfer{\Delta; \Gamma}{e : \tau}{\delta'}{} 
}{
  \typeinfer{\Delta; \Gamma}
   {\kwaction{a \oast (f := e)} : \typeaction{\delta \cup \delta'}{\omega \cup \{f\}}}
   {\emptyset}{}
}\and
\inferrule[]{
  \typeinfer{\Delta; \Gamma}{e : \typeaction{\delta'}{\omega}}{\delta}{} \\
  \dom{\delta'} = V_r, D_r
}{
  \reads{\Delta}{\kwdo{e}}{V_r, D_r}
}\and 
\end{mathpar}
Dynamics are defined w.r.t. runtime configuration 
$ \langle \Delta; S; Q_{r}; \\ Q_{do} \rangle $, 
where \(\Delta\) is typing environment calculated statically,
\(S\) maintains runtime information for all names, with the same domain as \(\Delta\) and partitioned into state variables \(V\) and definitions \(D\) where $f \mapsto \langle c, Prev\rangle \in V$ and $f \mapsto \langle c, e, Prev, Done,\\ Hist, Upda, Repi \rangle \in D$ to support strong consistency and glitch freedom proposed by Historiographer \cite{Freeman}.

Meerkat assumes developers and users don't care about the order of code updates or 
executing actions. Instead, they wait for result from last update or action to 
proceed on next. At runtime, two unordered sets
\(Q_{r} ::= \epsilon \mid Q_{r} \, \otimes \, r\) and \(Q_{do} ::= \epsilon \mid Q_{do} \, \otimes \, \kwdo{e}\) maintain pending updates/actions to process later. 
Following work by Costa Seco \cite{Seco}, \(\typeinfer{\Delta}{r : \Delta'}{}{}\) checks new typing 
environment contributed by \(r\). Notice for \(r_1 \oast r_2\), \(\Delta\) generated 
by \(r_1\) will effect \(r_2\) and latter one \(\Delta'\) can overwrite former one's environment by \(\Delta \uplus \Delta'\). As interaction from users,
\(\kwdo{e}\) should not alter \(\Delta\). \(\typcompat{\Delta}{ \Delta' }\) checks compatibility of 
\(\Delta\) followed by \(\Delta'\), i.e. well-formed (consistent and acyclic) of \(\Delta \uplus \Delta'\).
Now we describe rules for evolving code:
\begin{mathpar}\small
\inferrule[]{ }{
    \langle \Delta; S; Q_{r}; Q_{do} \rangle
    \stepact{\(\; r \;\)}
    \langle \Delta; S; Q_{r} \otimes r; Q_{do} \rangle
    }\and
\inferrule[]{
    \begin{tabular}{c}
    $\typeinfer{\Delta_{w_1}, \Delta'}{r_1: \Delta_1'}{}{}$ \;\;\;
    $\typeinfer{\Delta_{w_2}, \Delta'}{r_2: \Delta_2'}{}{} \;\;\;
    \typcompat{\Delta}{\Delta_1' \uplus \Delta_2'}$
    \end{tabular}\ 
 }{
    \langle \Delta = \Delta_{w_1}, \Delta_{w_2}, \Delta'; S; Q_{r} \otimes r_1 \otimes r_2;  Q_{do} \rangle
    \longrightarrow
    \langle \Delta \uplus \Delta_1' \uplus \Delta_2'; S; Q_{r}; Q_{do} \rangle
}\and
\inferrule[]{
    \typeinfer{\Delta}{r: \Delta'}{}{} \\
    \typcompat{\Delta}{\Delta'} \\ 
 }{
    \langle \Delta; S;  Q_{r} \otimes r; Q_{do} \rangle
    \longrightarrow
    \langle \Delta \uplus \Delta'; S; Q_{r}; Q_{do}  \rangle
    }\and
\inferrule[]{ }{
    \langle \Delta; S;  Q_{r}; Q_{do} \rangle
    \longrightarrow
    \langle \Delta; S; \epsilon; Q_{do}  \rangle
    }\and
\end{mathpar}

The system inserts \(r\) into set \(Q_r\) upon receiving the update from programmer, 
and later randomly pops out \(r\) and approves it to \(\Delta\), or concurrently pops out 
\(r_1\) and \(r_2\) without writing to same existing variables. We dynamically partition \(\Delta\) into write sets \(\Delta_{w_1}, \Delta_{w_2}\), and remaining \(\Delta'\).
The resource aware context tries to model waiting for locks when concurrent developers are updateing 
the code, where write locks allow both read and write, preventing other developers from reading the variable. Notice the premise \(\typeinfer{\Delta_{w_1}, \Delta'}{r_1: \Delta_1'}{}{}\) acquires all locks for \(r_1\), similar for \(r_2\). If there is no progress for acquiring locks even for one \(r\),
\(Q_r\) will ``die'' to empty and notify developers of rejected updates. 

Now we describe the rules for executing \(\kwdo{e}\)'s submitted by users (only present the rule for executing concurrently):
\begin{mathpar} \small
\inferrule[]{
\begin{tabular}{c}
$
S = V_{w_1},V_{w_2}, V_r, D  \;\;\;\;
\typeinfer{\Delta}{\kwdo{e_1} : \epsilon}{}{} \;\;\;\;
\typeinfer{\Delta}{\kwdo{e_2} : \epsilon}{}{} $\\
$\langle V_{w_1}; V_{r};  D \rangle \executes{\kwdo{e_1}}{} \langle V_{w_1}'; V_{r};  D' \rangle \;\;\; \langle V_{w_2}; V_{r};  D \rangle \executes{\kwdo{e_2}}{} \langle V_{w_2}'; V_{r};  D'' \rangle$ \\
$S' = V_{w_1'},V_{w_2'}, V_r, D' \cup D''$ \\
\end{tabular} 
    }{
    \langle \Delta; 
            S; 
            Q_{r};
            Q_{do} \otimes \kwdo{e_1} \otimes \kwdo{e_2}  \rangle 
    \longrightarrow 
    \langle S'; 
            Q_{r};
            Q_{do}
    \rangle 
}\and
\end{mathpar}

By splitting \(S\) into \(V_{w_1}, V_{w_2}\) the of state variables \(e_1\), \(e_2\) writes to respectively, state variables \(V_r\) (transitively) reads from, and \(D\) for all definition,
we propose similar acquirement for locks and Wait-Die mechanism as for evolving code, plus checking \(\kwdo{e}\) well-typed w.r.t. \(\Delta\). The stepping \(\executes{\kwdo{e}}{}\) abstracts 
Historiographer's\cite{Freeman} underlying propagation algorithm, updating the maps for \(V_w\) and \(D\)'s. \(D\) is allowed to share among concurrent executions thanks to confluent updates for \(D\) specified by Historiographer\cite{Freeman}. 

Finally, we propose theorems of type safety. The proof is still under construction and will 
be presented soon after.
\begin{theorem}[Runtime Progress]
   For all runtime configurations, \(\vdash \langle \Delta; S; Q_{r}; Q_{d o} \rangle\) implies 
   either \(Q_{r} = \epsilon \land Q_{do} = \epsilon \) or it steps to some configuration
   \( \langle \Delta'; S'; Q_{r}'; Q_{do}' \rangle\).
\end{theorem}
\begin{theorem}[Runtime Type Preservation]
   For all configurations \(\vdash \langle \Delta; S; Q_{r}; Q_{do} \rangle\) and \(\langle \Delta; S; Q_{r};\) \(Q_{do} \rangle \longrightarrow \langle \Delta'; S'; Q_{r}'; Q_{do}' \rangle\), then 
   \(\vdash \langle \Delta'; S'; Q_{r}'; Q_{do}' \rangle\).
\end{theorem}
\small\section*{Acknowledgements}
We thank Prof. Jonathan Aldrich and Prof. Jo\~{a}o Costa Seco for
their support and guidance on this work.

\printbibliography

\end{document}